\documentstyle[pra,aps,eqsecnum,preprint]{revtex}
\tightenlines
\begin{document}

\title{Noise-free scattering of the quantized electromagnetic field from a dispersive linear
dielectric}

\author{Mark Hillery \\
 Department of Physics \\
 Hunter College of the City University of New York \\
695 Park Ave, \\
 New York, NY 10021 USA\\
 Peter D. Drummond \\
 Department of Physics, The University of Queensland, \\
 Queensland 4072, Australia}

\maketitle
\begin{abstract}
We study the scattering of the quantized electromagnetic field from a linear,
dispersive dielectric using the scattering formalism for quantum fields. The
medium is modeled as a collection of harmonic oscillators with a number of distinct
resonance frequencies. This model corresponds to the Sellmeir expansion, which is widely used to describe experimental data for real dispersive media.
The integral equation for the interpolating field in
terms of the \( in \) field is solved and the solution used to find the \( out \)
field. The relation between the \( in \) and \( out \) creation and annihilation
operators is found which allows one to calculate the S-matrix for this system. In this
model, we find that there are absorption bands, but the input-output relations are
completely unitary. No additional quantum noise terms are required.
\end{abstract}

\section{Introduction}

A fundamental problem in quantum optics is how the properties of light change
as it propagates through a medium. If the medium is nonlinear, new frequencies
can be produced and the quantum noise properties of the field can be altered.
This leads to such interesting phenomena as solitons, squeezing, and quantum
phase diffusion, all of which have been observed \cite{nat}. If the medium
is linear, the situation is not as dramatic, but linear media serve as a first
step in the description of their nonlinear brethren, and they present problems
in their own right, such as the inclusion of dispersion. It is generally thought
that an accurate, first-principles treatment of dispersion necessitates the
inclusion of absorption, and consequently additional noise or reservoir operators. Here we analyse a quantum field theoretic model that demonstrates dispersion-induced absorption, but without any additional noise operators
appearing in the scattering relations.

Dielectric media can be described in a number of different ways. They can be
characterized by their susceptibilities, a procedure which, when fields are
included leads to the macroscopic Maxwell equations. Consequently, we shall
call this the macroscopic approach. It leads to difficulties when one wants
to include dispersion in a Lagrangian or Hamiltonian formulation, because dispersion
is a consequence of the fact that the response of the medium is not instantaneous,
but depends on the values of the field over a range of times \cite{landau}.
For certain kinds of fields, in particular narrow-band ones, these problems
can be overcome by using an approximate Lagrangian which is local in time \cite{drumm90}.
Another approach is to construct a microscopic model for the medium and to include
the degrees of freedom of the medium in the theory \cite{hop}-\cite{drumm98}.
This, for obvious reasons, we shall call the microscopic approach. It has the
advantage that the inclusion of dispersion is not a problem, but the disadvantage
is that for each new medium, a new model must be constructed. There are also
intermediate approaches which use frequency-dependent susceptibilities, but
also add quantum noise operators to the equations of motion for the fields \cite{loud},
\cite{grun}.

Most of the work on quantized fields in media has concentrated on what happens
inside the medium. However, there has been a steady stream of research which
has considered fields entering and leaving a medium as well. This is essential
if one wants to describe real experiments, in which the fields are generated
outside the medium, then pass through it, and are finally measured in free space.

Perhaps the first to investigate this question were Lang, et.\ al., who examined
the connection between the field inside and outside a laser cavity while studying
why the laser linewidth is so narrow \cite{lang1}, \cite{lang2}. Their model
consisted of a cavity bounded on one end by a perfectly reflecting wall and
on the other by a thin dielectric slab, and the cavity itself is filled with
an active medium. This cavity is embedded in a larger cavity which represents
the universe. They showed how the modes of the universe are related to the cavity
quasi-modes, which makes it possible to find the output field in terms of the
cavity field. In their analysis the field was classical, but soon thereafter
the quantum version of their model was constructed and used to investigate the
relation between the field inside and outside the cavity for a laser in the
linear regime by Ujihara \cite{uji}. This approach was later used by Gea-Banacloche,
et.\ al.\ to study the relationship between the squeezing generated inside a
cavity to that outside the cavity \cite{gea}.

This last problem had first been considered by Yurke who based his approach
on an earlier paper by Denker and himself \cite{yurke}. They developed a quantum
theory of electronic networks in which the network itself is located at \( x=0 \)
and transmission lines extending from there to \( x=+\infty  \) bring input
signals to the network and carry output signals away. Fields propagating toward
\( x=0 \) are input fields and those propagating away are output fields, and
the object is to find the output fields in terms of the input ones for a given
network. A related input-output theory was developed by Collett and Gardiner
\cite{gard} and was put on a firmer footing by Carmichael \cite{carm}. This
theory considers a cavity containing a medium, active or passive, linear or
nonlinear, which is coupled to a reservoir. The reservoir operators serve as
the input and output fields. The dynamics of the system inside the cavity is
described by a master equation, and its solution is used to find the time-dependent
reservoir operators and, thereby, the output field.

More recently a number of groups have examined the scattering of the quantized
electromagnetic field from inhomogeneous linear dielectrics. Glauber and Lewenstein
considered the case of a non-dispersive, lossless dielectric which is described
by a real position-dependent susceptibility \cite{glaub}.
The scattering from
dispersive media was studied by Kn\"{o}ll and Leonhardt who considered a medium
consisting of damped harmonic oscillators \cite{knoll92}. Rather than use a
formal scattering approach they used time-dependent Green's functions to solve
the field equations. A different treatment of dispersive media was given by
Matloob, et.\ al.\ \cite{loud}. They considered an arbitrary complex, frequency-dependent
dielectric function and quantized the theory at the level of the equations of
motion rather than starting with a Lagrangian. Working if frequency space they
found the fields emerging from a dielectric slab in terms of those entering
it. A similar analysis was carried out by Gruner and Welsch \cite{grun}. A
final approach is based on polaritons in finite media \cite{pol1}- \cite{pol3}.
For an infinite dielectric interacting with the electromagnetic field the eigenstates
of the Hamiltonian are mixed matter-field modes known as polaritons. If the
medium is finite, the polaritons acquire a finite lifetime. By looking at the
the electromagnetic parts of the polariton modes, scattering of the field from
the medium can be described.

In this paper we shall apply the quantum scattering formalism for fields to
describe an electromagnetic wave scattering from a finite medium, which behaves
as a mirror or beam-splitter. The medium is treated microscopically, and it
is dispersive. The final result is an explicit expression for the out operators
in terms of those of the input field. The calculation starts from first principles,
and, consequently, shows how some of the relations between in and out operators,
which are often used in quantum optics, follow from an underlying scattering
theory. An important feature of the model used here is that it includes a dielectric
medium with multiple bare resonances, leading to a number of discrete absorption
bands. This is typical of real dielectric materials, and leads to a dielectric
constant that can be modeled with the widely used Sellmeir\cite{BornWolf} expansion in frequency.

As we noted in our discussion of previous results, there are three treatments
of scattering from a dispersive, linear dielectric. Two are to some extent phenomenological
in that they do not start from a Hamiltonian describing the field-medium system
\cite{loud}, \cite{grun}.  The third approach
 treats time-dependent fields
rather than finding the asymptotic in and out fields which are the basic objects
in a scattering treatment \cite{knoll92}. 
We believe that this leaves room for a more fundamental
approach which can place the theory on a firmer foundation.
The results we find from field theoretic
scattering theory are similar to those found in the approach pioneered by Yurke,
and have the virtue that they are simple and intuitively clear. By employing
a fundamental approach, we have the advantage that the meaning of all of the
operators which we employ is well-defined, which is not always the case in the more
phenomenological treatments. The results presented here can be viewed as a justification
of earlier phenomenological theories.

\section{Model}

We shall consider a one-dimensional model of the electromagnetic field and the
medium which was developed in reference \cite{drumm98}. This model can be used
to describe the normal incidence of an electromagnetic wave on a medium, where
the wave travels in the \( x \) direction and is polarized in the \( z \)
direction.

The field can be represented by means of the dual potential, \( \Lambda (x,t) \),
which is appropriate if there are no free charges. In the case of a \( z \)-polarized
normally incident plane wave, \( \Lambda (x,t) \) is the \( y \) component
of the dual potential. The fields are given by \begin{equation}
D=\frac{\partial \Lambda }{\partial x}\hspace {1cm}B=\mu _{0}\frac{\partial \Lambda }{\partial t}\, \, .
\end{equation}

The medium consists of dipoles which are harmonic oscillators with masses \( m_{\nu } \)
and bare frequencies \( \Omega _{\nu } \), where \( \nu =1,\ldots ,N \). The un-renormalized
oscillator frequencies can be chosen to correspond to transition frequencies
of atoms or molecules making up an actual material. Each oscillator is described
by a field, \( r_{\nu }(x) \), which gives the displacement of the oscillator
at position \( x \) and with frequency \( \Omega _{\nu } \). It is convenient
to represent the oscillators in terms of the polarization fields \begin{equation}
p_{\nu }(x)=q_{\nu }\rho _{\nu }(x)r_{\nu }(x)\, \, ,
\end{equation}
 where \( \rho _{\nu }(x) \) is the density of oscillators with frequency \( \Omega _{\nu } \),
and the dipole corresponding to oscillators of type \( \nu  \) consists of
charges \( q_{\nu } \). We shall work in the multi-polar gauge so that
the coupling between the electromagnetic field and the medium is proportional
to \( \sum _{\nu }p_{\nu }(x)D(x) \). The medium self-interaction terms proportional
to the square of the total polarization are incorporated into the frequencies,
\( \Omega _{\nu } \).

For a volume such as a waveguide of cross-sectional area \( A \), the Lagrangian
density for the medium-field system is given by \begin{eqnarray}
\mathcal{L} & = & \frac{A}{2\epsilon _{0}}\left\{ \frac{1}{c^{2}}\dot{\Lambda }^{2}(x)-(\partial _{x}\Lambda (x))^{2}+\sum _{\nu }\left[ \frac{1}{g_{\nu }(x)}(\dot{p}_{\nu }^{2}(x)-\Omega _{\nu }^{2}p_{\nu }^{2}(x))\right. \right. \nonumber \\
 &  & \left. \left. +2p_{\nu }(x)\partial _{x}\Lambda (x)\right] \right\} \, \, ,
\end{eqnarray}
 where \begin{equation}
g_{\nu }(x)=\frac{q_{\nu }^{2}\rho _{\nu }(x)}{m_{\nu }\epsilon _{0}}\, \, .
\end{equation}

\subsection{Refractive index}

>From the Lagrangian density we find the equations of motion for the fields \begin{eqnarray}
\partial _{t}^{2}\Lambda -c^{2}\partial _{x}^{2}\Lambda  & = & -c^{2}\partial _{x}\sum _{\nu }p_{\nu }\nonumber \\
\partial _{t}^{2}p_{\nu }+\Omega _{\nu }^{2}p_{\nu } & = & g_{\nu }\partial _{x}\Lambda \, \, .\label{mode-equn} 
\end{eqnarray}
 For a medium of constant density, (i.\ e.\, \( g_{\nu }(x) \) is independent
of \( x \)), we can solve the above equations by assuming that both \( p_{\nu } \)
and \( \Lambda  \) are proportional to \( e^{i(kx-\omega t)} \). The values
of \( \omega  \) are the frequencies of the modes of the system and are given
by the solutions of equation {[}\ref{mode-equn}{]}: \begin{equation}
\omega ^{2}=(kc)^{2}\left[ 1-\sum _{\nu }\frac{g_{\nu }}{\Omega _{\nu }^{2}-\omega ^{2}}\right] \, \, .
\end{equation}
 Defining the index of refraction, \( n(\omega ) \), to be \( kc/\omega  \),
we find \begin{equation}
n(\omega )=\left[ 1-\sum _{\nu }\frac{g_{\nu }}{\Omega _{\nu }^{2}-\omega ^{2}}\right] ^{-1/2}.
\end{equation}
 This is very similar\cite{drumm98} to the classical Sellmeir expansion for
the refractive index. Note that this expansion is not identical to the Sellmeir
expansion, but can be converted into the commonly used Sellmeir form through
a renormalization of the bare resonant frequencies of the oscillators. The characteristic
property of this type of equation is that it possesses solutions for the refractive
index that are either purely real (transmission bands) or purely imaginary (absorption
bands). At the bare resonance frequency, the refractive index is zero. Near
a resonance, where \( \omega \rightarrow \Omega _{\nu } \), the refractive
index is real for \( \omega >\Omega _{\nu } \), and imaginary for \( \omega <\Omega _{\nu } \).
At a finite detuning below a resonance, the refractive index goes to infinity
just below the start of the corresponding absorption band.

\subsection{Lagrangian quantization}

>From the Lagrangian density we can find the canonical momenta corresponding
to \( \Lambda  \) and \( p_{\nu } \), which we shall denote by \( \Pi  \)
and \( \pi _{\nu } \), respectively. These are given by \begin{equation}
\Pi (x)=\mu _{0}\dot{\Lambda }(x)\hspace {1cm}\pi _{\nu }(x)=\frac{\dot{p}_{\nu }(x)}{\epsilon _{0}g_{\nu }(x)}\, \, .
\end{equation}
 The theory is quantized by imposing the commutation relations \begin{equation}
[\hat{\Lambda }(x,t),\hat{\Pi }(x^{\prime },t)]=i\hbar \delta (x-x^{\prime })/A
\end{equation}
 and \begin{equation}
[\hat{p}_{\nu }(x,t),\hat{\pi }_{\nu ^{\prime }}(x^{\prime },t)]=i\hbar \delta _{\nu ,\nu ^{\prime }}\delta (x-x^{\prime })/A\, \, .
\end{equation}
 The canonical momenta and the Lagrangian density can now be used to find the
Hamiltonian density for the quantized theory 
\begin{eqnarray}
{\mathcal H}(x) & = & \frac{A}{2\epsilon _{0}}:\left\{ \frac{\epsilon _{0}}{\mu _{0}}\hat{\Pi }^{2}(x)+(\partial _{x}\hat{\Lambda }(x))^{2}+\sum _{\nu }[\epsilon _{0}^{2}g_{\nu }(x)\hat{\pi }_{\nu }^{2}(x)\right. \nonumber \\
 &  & \left. +\frac{\Omega _{\nu }^{2}}{g_{\nu }(x)}\hat{p}_{\nu }^{2}(x)-2\hat{p}_{\nu }(x)\partial _{x}\hat{\Lambda }(x)]\right\} :\, \, .
\end{eqnarray}
 This can be put in a different form if we define annihilation and creation
operators, \( \hat{\xi }_{\nu }(x) \) and \( \hat{\xi }_{\nu }^{\dagger }(x) \),
for the oscillators, where \begin{equation}
\hat{\xi }_{\nu }(x)=\frac{1}{\sqrt{2\hbar }}\left( \sqrt{\frac{\Omega _{\nu }}{\epsilon _{0}g_{\nu }(x)}}\hat{p}_{\nu }+i\sqrt{\frac{\epsilon _{0}g_{\nu }(x)}{\Omega _{\nu }}}\hat{\pi }_{\nu }\right) ,
\end{equation}
 so that \begin{equation}
[\hat{\xi }_{\nu }(x),\hat{\xi }_{\nu ^{\prime }}^{\dagger }(x^{\prime })]=\delta _{\nu ,\nu ^{\prime }}\delta (x-x^{\prime })/A.
\end{equation}
 We finally have for the Hamiltonian density \begin{eqnarray}
{\mathcal H}(x) & = & \frac{A}{2\epsilon _{0}}:\left\{ \frac{\epsilon _{0}}{\mu _{0}}\hat{\Pi }^{2}(x)+(\partial _{x}\hat{\Lambda }(x))^{2}+\sum _{\nu }[2\epsilon _{0}\hbar \Omega _{\nu }\hat{\xi }_{\nu }^{\dagger }(x)\hat{\xi }_{\nu }(x)\right. \nonumber \\
 &  & \left. -2\sqrt{\frac{\hbar \epsilon _{0}g_{\nu }(x)}{2\Omega _{\nu }}}(\hat{\xi }_{\nu }(x)+\hat{\xi }_{\nu }^{\dagger }(x))\partial _{x}\hat{\Lambda }(x)]\right\} :
\end{eqnarray}

\section{Scattering Theory}

In order to determine what happens when an electromagnetic wave scatters off
of the medium we shall apply the standard formulation of scattering for quantum
fields \cite{hen}. This is done in the Heisenberg picture so that it is the
field operators which are time dependent. Because we shall consider a medium
which is bounded in the \( x \) direction, the interaction is bounded in time.
This can be seen either by considering the incoming waves to be wave packets,
so that the interaction takes place only while the packet is inside the medium,
or by using plane waves and turning the interaction on and off adiabatically.
In either approach, the fields will go to free fields both as \( t\rightarrow -\infty  \)
and as \( t\rightarrow \infty  \). The free fields as \( t\rightarrow -\infty  \)
are the in fields, and those as \( t\rightarrow \infty  \) are the out fields.
The time dependent field operators which carry the full time dependence of the
Hamiltonian, including the interaction, are also known as the interpolating
fields, because they interpolate between the in and the out fields. Our goal
is to use the interpolating fields to find an expression for the out fields
in terms of the in fields. This will give us a complete description of the scattering
process.We note here that a related Heisenberg-picture approach to quantum scattering theory relevant to quantum optics measurements was recently developed by Dalton et al.\cite{dalton}. They developed a similar basic formalism, but did
not consider specific examples.

\subsection{\emph{In} and \emph{out} fields}

To find the relationship between the in and out-fields, we need to express the
equations of motion of the interpolating fields as integral equations. From
the Hamiltonian for our model we find \begin{eqnarray}
(\partial _{t}^{2}-c^{2}\partial _{x}^{2})\hat{\Lambda } & = & -c^{2}\partial _{x}\sum _{\nu }\sqrt{\frac{\hbar \epsilon _{0}g_{\nu }}{2\Omega _{\nu }}}\left[ \hat{\xi }_{\nu }+\hat{\xi }_{\nu }^{\dagger }\right] \nonumber  \\
(\partial _{t}+i\Omega _{\nu })\hat{\xi }_{\nu } & = & \frac{i}{\epsilon _{0}\hbar }\sqrt{\frac{\hbar \epsilon _{0}g_{\nu }}{2\Omega _{\nu }}}\partial _{x}\hat{\Lambda }\, \, .\label{flddiff}
\end{eqnarray}
 In order to express these as integral equations we define the Green's functions
\( \Delta ^{(ret)}(x,t) \), \( \Delta ^{(adv)}(x,t) \), \( \Gamma ^{(ret)}_{\nu }(x,t) \),
and \( \Gamma ^{(adv)}_{\nu }(x,t) \). They satisfy the equations \begin{eqnarray}
(\partial _{t}^{2}-c^{2}\partial _{x}^{2})\Delta ^{(ret)}(x,t) & = & \delta (x)\delta (t)\nonumber \\
(\partial _{t}+i\Omega _{\nu })\Gamma ^{(ret)}_{\nu }(x,t) & = & \delta (x)\delta (t)\nonumber \\
(\partial _{t}^{2}-c^{2}\partial _{x}^{2})\Delta ^{(adv)}(x,t) & = & \delta (x)\delta (t)\nonumber \\
(\partial _{t}+i\Omega _{\nu })\Gamma ^{(ret)}_{\nu }(x,t) & = & \delta (x)\delta (t)\, \, ,
\end{eqnarray}
 and the boundary conditions \begin{eqnarray}
\Delta ^{(ret)}(x,t)=\Gamma ^{(ret)}_{\nu }(x,t)&=&0 \, \, \, {\textrm{for}} \, t<0\nonumber \\
\Delta ^{(adv)}(x,t)=\Gamma ^{(adv)}_{\nu }(x,t)&=&0 \,\, \, {\textrm{for}} \, t>0\, \, .
\end{eqnarray}
 The retarded Green's functions can be expressed as \begin{eqnarray}
\Delta ^{(ret)}(x,t) & = & \frac{1}{(2\pi )^{2}}\int dkd\omega \frac{e^{i(kx-\omega t)}}{(ck)^{2}-(\omega +i\epsilon )^{2}}\nonumber \\
\Gamma ^{(ret)}_{\nu }(x,t) & = & \frac{\delta (x)}{2\pi }\int d\omega \frac{e^{-i\omega t}}{i(\Omega -i\epsilon -\omega )}\, \, ,
\end{eqnarray}
 where \( \epsilon \rightarrow 0^{+} \), and the advanced Green's functions
are given by almost identical expressions, the only difference being that \( \epsilon  \)
is replaced by \( -\epsilon  \). The integral equations corresponding to the
differential equations, Eqs. (\ref{flddiff}), are \begin{eqnarray}
\hat{\Lambda }(x,t) & = & \hat{\Lambda }_{in}(x,t)-c^{2}\int dx^{\prime }\int dt^{\prime }\Delta ^{(ret)}(x-x^{\prime },t-t^{\prime })\nonumber  \\
 &  & \partial _{x^{\prime }}\sum _{\nu }\sqrt{\frac{\hbar \epsilon _{0}g_{\nu }(x^{\prime })}{2\Omega _{\nu }}}\left[ \hat{\xi }_{\nu }(x^{\prime },t)+\hat{\xi }_{\nu }^{\dagger }(x^{\prime },t)\right] \nonumber \\
\hat{\xi }_{\nu }(x,t) & = & \hat{\xi }^{(in)}_{\nu }(x,t)+\frac{i}{\epsilon _{0}\hbar }\int dx^{\prime }\int dt^{\prime }\Gamma ^{(ret)}(x-x^{\prime },t-t^{\prime })\nonumber \\
 &  & \sqrt{\frac{\hbar \epsilon _{0}g_{\nu }(x^{\prime })}{2\Omega _{\nu }}}\partial _{x^{\prime }}\hat{\Lambda }(x^{\prime },t^{\prime })\, \, .
 \label{intret}
\end{eqnarray}
 The corresponding expression involving the out-fields is: \begin{eqnarray}
\hat{\Lambda }(x,t) & = & \hat{\Lambda }_{out}(x,t)-c^{2}\int dx^{\prime }\int dt^{\prime }\Delta ^{(adv)}(x-x^{\prime },t-t^{\prime })\nonumber  \\
 &  & \partial _{x^{\prime }}\sum _{\nu }\sqrt{\frac{\hbar \epsilon _{0}g_{\nu }(x^{\prime })}{2\Omega _{\nu }}}\left[ \hat{\xi }_{\nu }(x^{\prime },t)+\hat{\xi }_{\nu }^{\dagger }(x^{\prime },t)\right] \nonumber \\
\hat{\xi }_{\nu }(x,t) & = & \hat{\xi }^{(out)}_{\nu }(x,t)+\frac{i}{\epsilon _{0}\hbar }\int dx^{\prime }\int dt^{\prime }\Gamma ^{(adv)}(x-x^{\prime },t-t^{\prime })\nonumber \\
 &  & \sqrt{\frac{\hbar \epsilon _{0}g_{\nu }(x^{\prime })}{2\Omega _{\nu }}}\partial _{x^{\prime }}\hat{\Lambda }(x^{\prime },t^{\prime })\, \, .
 \label{intadv}
\end{eqnarray}
 Note that the integral equations incorporate the boundary conditions for the
fields. The first set of equations implies that \( \hat{\Lambda }(x,t) \) and
\( \hat{\xi }_{\nu }(x,t) \) will go to \( \hat{\Lambda }_{in}(x,t) \) and
\( \hat{\xi }^{(in)}_{\nu }(x,t) \), respectively, as \( t\rightarrow -\infty  \),
and the second set implies that they will go to \( \hat{\Lambda }_{out}(x,t) \)
and \( \hat{\xi }^{(out)}_{\nu }(x,t) \), respectively, as \( t\rightarrow \infty  \). 

What we shall do is solve the first set of equations for \( \hat{\Lambda }(x,t) \)
and \( \hat{\xi }_{\nu }(x,t) \) in terms of the in fields, and then insert
this solution into the second set to find the out fields in terms of the in
fields.

\subsection{Fourier decomposition}

We begin solving Eqs. (\ref{intret}) by taking the time Fourier transform of
both sides. Defining \begin{eqnarray}
\hat{\Lambda }(x,\omega ) & = & \frac{1}{\sqrt{2\pi }}\int dte^{i\omega t}\hat{\Lambda }(x,t)\nonumber \\
\hat{\xi }_{\nu }(x,\omega ) & = & \frac{1}{\sqrt{2\pi }}\int dte^{i\omega t}\hat{\xi }_{\nu }(x,t)\, \, ,
\end{eqnarray}
 and similarly for the in and out-fields, we find that \begin{eqnarray}
\hat{\Lambda }(x,\omega ) & = & \hat{\Lambda }_{in}(x,\omega )-\frac{ic}{2\omega }\int dx^{\prime }e^{i\omega |x-x^{\prime }|/c}\partial _{x^{\prime }}\sum _{\nu }\sqrt{\frac{\hbar \epsilon _{0}g_{\nu }(x^{\prime })}{2\Omega _{\nu }}}\nonumber  \\
 &  & \hspace {2cm}\left[ \hat{\xi }_{\nu }(x^{\prime },\omega )+\hat{\xi }_{\nu }^{\dagger }(x^{\prime },-\omega )\right] \, \, ,
 \label{gamk}
\end{eqnarray}
 and \begin{equation}
\label{xik}
\hat{\xi }_{\nu }(x,\omega )=\hat{\xi }^{(in)}_{\nu }(x,\omega )+\frac{1}{\epsilon _{0}\hbar }\sqrt{\frac{\hbar \epsilon _{0}g_{\nu }(x)}{2\Omega _{\nu }}}\frac{1}{\Omega _{\nu }-i\epsilon -\omega }\partial _{x}\hat{\Lambda }(x,\omega )\, \, .
\end{equation}
 In deriving these equations we made use of the fact that \begin{equation}
\int dte^{i\omega t}\Delta ^{(ret)}(x,t)=\frac{i}{2\omega c}e^{i\omega |x|/c}\, \, .
\end{equation}
 We can derive an equation for only the field \( \hat{\Lambda }(x,k_{0}) \)
by substituting from Eq. (\ref{xik}) into Eq. (\ref{gamk}). We find that \begin{eqnarray}
\hat{\Lambda }(x,\omega ) & = & \hat{\Lambda }_{in}(x,\omega )-\frac{ic}{2\omega }\int dx^{\prime }e^{i\omega |x-x^{\prime }|/c}\partial _{x^{\prime }}\sum _{\nu }\sqrt{\frac{\hbar \epsilon _{0}g_{\nu }(x^{\prime })}{2\Omega _{\nu }}}\nonumber \\
 &  & \left[ \hat{\xi }_{\nu }^{(in)}(x^{\prime },\omega )+\hat{\xi }_{\nu }^{(in)\dagger }(x^{\prime },-\omega )\right. \nonumber \\
 &  & \left. +\frac{1}{\epsilon _{0}\hbar }\sqrt{\frac{\hbar \epsilon _{0}g_{\nu }(x^{\prime })}{2\Omega _{\nu }}}\frac{2\Omega _{\nu }}{\Omega _{\nu }^{2}-(\omega +i\epsilon )^{2}}\partial _{x^{\prime }}\hat{\Lambda }(x^{\prime },\omega )\right] \, \, .\label{solution} 
\end{eqnarray}

Our next step is to turn this into a differential equation, but before doing
so we shall make a simplifying assumption. The field \( \hat{\xi }_{\nu }^{(in)}(x,t) \)
is a free field which oscillates at the frequency \( \Omega _{\nu } \), and
this implies that \( \hat{\xi }_{\nu }^{(in)}(x,\omega ) \) is nonzero only
when \( \omega =\Omega _{\nu } \) . We are mainly interested in cases where
the incoming light is not resonant with the medium, so we shall initially assume
that \( \omega \neq \Omega _{\nu } \) for \( \nu =1,\ldots N \). This implies
that we can drop \( \hat{\xi }_{\nu }^{(in)}(x,\omega ) \) and \( \hat{\xi }_{\nu }^{(in)\dagger }(x,-\omega ) \)
from the above equation and set \( \epsilon =0 \). We return to the resonant
case later.

Next, we then apply the differential operator \( c^{2}\partial _{x}^{2}+\omega ^{2} \)
to both sides. This annihilates the in-field term and converts the integral
equation into a homogeneous differential equation. The result is \begin{equation}
\label{kdiff}
\partial _{x}\left( \frac{c^{2}}{n^{2}(x,\omega )}\partial _{x}\hat{\Lambda }(x,\omega )\right) +\omega ^{2}\hat{\Lambda }(x,\omega )=0\, \, .
\end{equation}
 Here, \( n(x,\omega ) \), the space and frequency dependent index of refraction
of the medium, is given by \begin{equation}
n(x,\omega )=\left( 1-\sum _{\nu }\frac{g_{\nu }(x)}{\Omega _{\nu }^{2}-\omega ^{2}}\right) ^{-1/2}\, \, .
\end{equation}

In this form, the equations have a rather classical appearance, and the matter
operators no longer appear in the formulation, which gives rise to a substantial
simplification.

\section{Dielectric layer}

We now want to specialize our equations to the case of a dielectric layer with
a uniform density of oscillators. This corresponds to the important case of
a beam-splitter or mirror, although we make no restrictions as to the size of
the layer. The medium extends from \( x=-L \) to \( x=L \). Inside the medium,
\( n(x,\omega ) \) has a value of \( n_{0}(\omega ) \), and outside the medium
it has a value of \( 1 \). The solutions to Eq. (\ref{kdiff}) should be continuous
and \( n^{-2}(x,\omega ) \) times their derivative should be continuous. These
correspond to a continuous magnetic and electric field, respectively.

\subsection{Classical Case}

In order to find the solution of the operator equation, Eq. (\ref{kdiff}),
we first find solutions of the corresponding c-number equation, which we shall
denote as \( u(x,\omega ) \). We begin by dividing the line into three regions,
region I for \( x<-L \), region II for \( -L\leq x\leq L \), and region III
for \( x>L \). In regions I and III, \( u(x,\omega ) \) satisfies \begin{equation}
\label{diff13}
\left[ c^{2}\partial _{x}^{2}+\omega ^{2}\right] u(x,\omega )=0\, \, ,
\end{equation}
 and in region II \begin{equation}
\label{diff2}
\left[ c^{2}\partial _{x}^{2}+\omega ^{2}n_{0}^{2}(\omega )\right] u(x,\omega )=0\, \, ,
\end{equation}
 where \( n_{0} \) is the value of \( n(x,\omega ) \) in region II. A solution
incident from the left, \( u_{l}(x,\omega ) \), which satisfies the equation
and has the proper continuity properties, is given by \begin{equation}
u_{l}(x,\omega )=\left\{ \begin{array}{cl}
e^{ik(\omega )x}+R(\omega )e^{-ik(\omega )x} & \mbox {in\, \, region\, \, I}\\
B^{(l)}_{r}(\omega )e^{i\kappa (\omega )x}+B^{(l)}_{l}(\omega )e^{-i\kappa (\omega )x} & \mbox {in\, \, region\, \, II}\\
T(\omega )e^{ik(\omega )x} & \mbox {in\, \, region\, \, III\, \, ,}
\end{array}\right. 
\end{equation}
 where \( k(\omega )=\omega /c \) , and \( \kappa (\omega )=n_{0}k(\omega ) \)
. It is to be remembered that waves proportional to \( e^{ikx} \) or \( e^{i\kappa x} \)
are propagating to the right, and those proportional to \( e^{-ikx} \) or \( e^{-i\kappa x} \)
are propagating to the left. 

Suppressing the frequency arguments for clarity, the coefficients in the above
equation are given by \begin{eqnarray}
R=-\frac{i(n_{0}^{2}-1)\sin (2\kappa L)}{D}e^{-2ikL} & \hspace{1cm} 
& T=\frac{2n_{0}}{D}e^{-2ikL}\nonumber \\
B^{(l)}_{r}=\frac{n_{0}(n_{0}+1)}{D}e^{-i(\kappa +k)L} & \hspace{1cm} 
& B^{(l)}_{l}=-\frac{n_{0}(n_{0}-1)}{D}e^{i(\kappa -k)L}\, \, ,
\end{eqnarray}
where \( D=2n_{0}\cos (2\kappa L)-i(n_{0}^{2}+1)\sin (2\kappa L) \). Note
that \( R \) and \( T \) are, respectively, the reflection and transmission
coefficients for the medium, and that \( |R|^{2}+|T|^{2}=1 \). For a solution
incident from the right we have \begin{equation}
u_{r}(x,\omega )=\left\{ \begin{array}{cl}
T(\omega )e^{-ik(\omega )x} & \mbox {in\, \, region\, \, I}\\
B^{(r)}_{r}(\omega )e^{i\kappa (\omega )x}+B^{(r)}_{l}(\omega )e^{-i\kappa (\omega )x} & \mbox {in\, \, region\, \, II}\\
e^{-ik(\omega )x}+R(\omega )e^{ik(\omega )x} & \mbox {in\, \, region\, \, III\, \, ,}
\end{array}\right. 
\end{equation}
 where \( k \) , \( \kappa  \), \( R \) and \( T \) are as before, while
\( B^{(r)}_{r}=B^{(l)}_{l} \), and \( B^{(r)}_{l}=B^{(l)}_{r} \).

\subsection{Asymptotic fields}

Both \( u_{r} \) and \( u_{l} \) are solutions of the differential equation,
Eqs. (\ref{diff13}) and (\ref{diff2}), and this implies that they are also
solutions to the corresponding integral equation \begin{eqnarray}
\Lambda (x,\omega )= & \Lambda _{in}(x,\omega ) & -\frac{ic}{2\omega }\int _{-L}^{L}dx^{\prime }e^{i\omega |x-x^{\prime }|/c}\partial _{x^{\prime }}\nonumber  \\
 &  & \left[ \left( 1-\frac{1}{n^{2}(x,\omega )}\right) \partial _{x^{\prime }}\Lambda (x^{\prime },\omega )\right] \, \, ,
 \label{kint}
\end{eqnarray}
 for particular choices of the field \( \Lambda _{in}(x,\omega ) \). We can
find \( \Lambda _{in}(x,\omega ) \) for both solutions simply by substituting
them into Eq. (\ref{kint}). We must be careful, however, because the expression
inside the square brackets is not continuous at \( x=\pm L \), and it is being
differentiated, so that the discontinuities will lead to finite contributions
after being integrated. One way to find these contributions is to consider a
refractive index which is continuous, but which goes to the desired one as a
limit. 

For example, let us suppose that \( n(x,\omega ) \) is \( 1 \) for \( x<-L-\delta  \)
and \( x>L+\delta  \), is equal to \( n_{0} \) for \( -L\leq x\leq L \),
goes continuously from \( 1 \) to \( n_{0} \) as \( x \) goes from \( -L-\delta  \)
to \( -L \), and goes continuously from \( n_{0} \) to \( 1 \) as \( x \)
goes from \( L \) to \( L+\delta  \). We can then take the limit \( \delta \rightarrow 0 \).
Let us examine what happens in the interval between \( -L-\delta  \) and \( -L \);
the interval between \( L \) and \( L+\delta  \) is similar. As \( \delta \rightarrow 0 \)
we have that \begin{eqnarray}
\int _{-L-\delta }^{-L}dx^{\prime }e^{i\omega |x-x^{\prime }|/c}\partial _{x^{\prime }}\left[ \left( 1-\frac{1}{n^{2}(x,\omega )}\right) \partial _{x^{\prime }}\Lambda (x^{\prime },\omega )\right]  &  & \nonumber \\
\rightarrow e^{i\omega |x+L|/c}\int _{-L-\delta }^{-L}dx^{\prime }\partial _{x^{\prime }}\left[ \left( 1-\frac{1}{n^{2}(x,\omega )}\right) \partial _{x^{\prime }}\Lambda (x^{\prime },\omega )\right]  &  & \nonumber \\
=\left. e^{i\omega |x+L|/c}\left( 1-\frac{1}{n^{2}_{0}(\omega )}\right) \partial _{x}\Lambda (x,\omega )\right| _{x=-L^{+}}\, \, , & 
\end{eqnarray}
 where \( x=-L^{+} \) denotes the limit as \( x\rightarrow -L \) from the
positive direction (\( x=L^{-} \) is defined in an analogous fashion). The
other limit of the integral contributes zero, due to the refractive index term
approaching unity. Explicitly putting in the terms resulting from the boundaries
of the medium gives \begin{eqnarray}
\Lambda (x,\omega )= & \Lambda _{in}(x,\omega ) & -\frac{ic}{2\omega }\int _{-L^{+}}^{L^{-}}dx^{\prime }e^{i\omega |x-x^{\prime }|/c}\left( 1-\frac{1}{n^{2}_{0}(\omega )}\right) \partial _{x^{\prime }}^{2}\Lambda (x^{\prime },\omega )\nonumber  \\
 &  & -\frac{ic}{2\omega }\left[ e^{i\omega |x+L|/c}\left( 1-\frac{1}{n^{2}_{0}(\omega )}\right) \partial _{x}\left. \Lambda (x,\omega )\right| _{x=-L^{+}}\right. \nonumber \\
 &  & \left. -e^{i\omega |x-L|/c}\left( 1-\frac{1}{n^{2}_{0}(\omega )}\right) \partial _{x}\left. \Lambda (x,\omega )\right| _{x=-L^{-}}\right] \, \, .
 \label{kintb}
\end{eqnarray}
 If we now substitute \( u_{l}(x,\omega ) \) into this equation instead of
\( \Lambda (x,\omega ) \), we find that \begin{equation}
\Lambda _{in}(x,\omega )=e^{i\omega x/c}\, \, ,
\end{equation}
 and if we substitute \( u_{r}(x,\omega ) \), we find \begin{equation}
\Lambda _{in}(x,\omega )=e^{-i\omega x/c}\, \, .
\end{equation}

\subsection{Quantum Case}

In the quantum case we have the usual expansion of a free field in terms of
annihilation and creation operators. This leads, in the present case, to: \begin{equation}
\hat{\Lambda }_{in}(x,t)=\int dk\sqrt{\frac{\hbar c\epsilon _{0}}{4\pi A|k|}}\left[ \hat{a}^{(in)}_{k}e^{i(kx-|k|ct)}+(\hat{a}^{(in)}_{k})^{\dagger }e^{-i(kx-|k|ct)}\right] \, \, ,
\end{equation}
 which implies that for \( \omega >0 \)\begin{equation}
\label{gamin}
\hat{\Lambda }_{in}(x,\omega )=\sqrt{\frac{\hbar \epsilon _{0}}{2cAk(\omega )}}
\left[ \hat{a}^{(in)}_{k(\omega )}e^{i\omega x/c}+
\hat{a}^{(in)}_{-k(\omega )}e^{-i\omega x/c}\right] \, \, .
\end{equation}
 The results of the previous paragraph allow us to see that if \( \hat{\Lambda }(x,\omega ) \)
is given by \begin{equation}
\label{interp}
\hat{\Lambda }(x,\omega )=\sqrt{\frac{\hbar \epsilon _{0}}{2cAk(\omega )}}\left[ \hat{a}^{(in)}_{k(\omega )}u_{l}(x,\omega )+\hat{a}^{(in)}_{-k(\omega )})u_{r}(x,\omega )\right] \, \, ,
\end{equation}
 then it is a solution of Eq. (\ref{kintb}) with \( \hat{\Lambda }_{in} \)
given by Eq. (\ref{gamin}). This gives us the interpolating field in terms
of the in field.

Our remaining task is to use the expression for the interpolating field to find
the out-field in terms of the in-field. This can be done by substituting the
expression for \( \hat{\Lambda }(x,\omega ) \) given in the previous paragraph
into the equation which relates the interpolating field to the out-field \begin{eqnarray}
\hat{\Lambda }(x,\omega )= & \hat{\Lambda }_{out}(x,\omega ) & +\frac{ic}{2\omega }\int _{-L^{+}}^{L^{-}}dx^{\prime }e^{-i\omega |x-x^{\prime }|/c}\left( 1-\frac{1}{n^{2}_{0}(\omega )}\right) \partial _{x^{\prime }}^{2}\hat{\Lambda }(x^{\prime },\omega )\nonumber \\
 &  & +\frac{ic}{2\omega }\left[ e^{-i\omega |x+L|/c}\left( 1-\frac{1}{n^{2}_{0}(\omega )}\right) \partial _{x}\left. \hat{\Lambda }(x,\omega )\right| _{x=-L^{+}}\right. \nonumber \\
 &  & \left. -e^{-i\omega |x-L|/c}\left( 1-\frac{1}{n^{2}_{0}(\omega )}\right) \partial _{x}\left. \hat{\Lambda }(x,\omega )\right| _{x=-L^{-}}\right] \, \, ,
\end{eqnarray}
 which follows from Eqs. (\ref{intadv}). The derivation is almost identical
to that of Eq. (\ref{kintb}), so we do not give it explicitly. Making this
substitution we find that, for \( \omega >0 \)\begin{eqnarray}
\hat{\Lambda }_{out}(x,\omega ) & = & \sqrt{\frac{\hbar \epsilon _{0}}{2cAk(\omega )}}\left[ (T(\omega )e^{ik(\omega )x}+R(\omega )e^{-ik(\omega )x})\hat{a}^{(out)}_{k(\omega )}\right. \nonumber  \\
 &  & \left. +(R(\omega )e^{ik(\omega )x}+T(\omega )e^{-ik(\omega )x})\hat{a}^{(out)}_{-k(\omega )}\right] \, \, .
 \label{kout1}
\end{eqnarray}

The out-field can also be expressed in terms of out creation and annihilation
operators, \begin{equation}
\hat{\Lambda }_{out}(x,t)=
\int dk\sqrt{\frac{\hbar c\epsilon _{0}}{4\pi A|k|}}\left[ \hat{a}^{(out)}_{k}e^{i(kx-|k|ct)}+(\hat{a}^{(out)}_{k})^{\dagger }e^{-i(kx-|k|ct)}\right] \, \, .
\end{equation}
 Taking the Fourier transform of this equation with respect to time, for \( \omega >0 \),
gives \begin{equation}
\label{kout2}
\hat{\Lambda }_{out}(x,\omega )=\sqrt{\frac{\hbar \epsilon _{0}}{2cAk(\omega )}}\left[ \hat{a}^{(out)}_{k(\omega )}e^{ik(\omega )x}+\hat{a}^{(out)}_{-k(\omega )}e^{-ik(\omega )x})\right] \, \, .
\end{equation}
 Comparing Eqs. (\ref{kout1}) and (\ref{kout2}) we see that for \( \omega >0 \)\begin{eqnarray}
\hat{a}_{k(\omega )}^{(out)} & = & T(\omega )\hat{a}_{k(\omega )}^{(in)}+R(\omega )\hat{a}_{-k(\omega )}^{(in)}\\
\hat{a}_{-k(\omega )}^{(out)} & = & 
R(\omega )\hat{a}_{k(\omega )}^{(in)}+T(\omega )\hat{a}_{-k(\omega )\, \, }^{(in)}
\, \, . \label{inout}
\end{eqnarray}

These equations are the solution to the scattering problem. We note that the
transmission and reflection coefficients satisfy the usual relation of
 \( |T(\omega )|^{2}+|R(\omega )|^{2}=1 \).
This holds even in the bandgap regions, where transmission occurs via an evanescent
field. Thus, the scattering problem is explicitly unitary, as we would expect.
It is important to notice that unitarity holds even inside the band-gaps of
the problem, indicating that the absorption bands simply modify the reflection
and transmission coefficients, without removing photons. Another way to think
of this, is that even when a photon is removed through virtual excitation of
an atomic resonance, the photon will eventually be re-radiated in either the
forward of backward directions.

\subsection{Resonances}

Let us now consider what happens when \( \omega =\pm \Omega _{\nu }. \) This
case was excluded from our earlier treatment, since it must involve some matter-operator
contribution, which we have neglected so far. Of course, as the resonances are
discrete, the frequencies involved are essentially a set of measure zero, lying
on the upper edge of each band-gap.

In this case, terms proportional to either \( \hat{\xi }_{\nu }^{(in)}(x,\omega ) \)
or \( \hat{\xi }_{\nu }^{(in)\dagger }(x,-\omega ) \) will be present in Eq.
(\ref{solution}). This, in turn, means that the solution for the interpolating
field given in Eq.(\ref{interp}) must be modified. In particular, a term proportional
to the matter fields must be added. This is not surprising; it is usually accepted
that in a dispersive medium there must be absorption, and this in turn generally
requires coupling to a reservoir field. However, it would be rather surprising
to find quantum noise only occurring at discrete frequencies corresponding to
the band edge. We have already established, in particular, that absorption in
the band at frequencies different from the resonances simply changes the transmission
and reflection coefficients, without adding any noise source. We will now show
that even with the matter terms included, our earlier conclusions still hold;
there are no extra noise sources for the out-fields.

In order to see this, we first note that 
\begin{equation}
\hat{\xi }_{\nu }^{(in)}(x,t)=
e^{-i\Omega _{\nu }t}\, \hat{\xi }_{\nu }^{(in)}(x)\, \, ,
\end{equation}

where: \( \hat{\xi }_{\nu }^{(in)}(x)=e^{-i\Omega _{\nu }t}\, \hat{\xi }_{\nu }^{(in)}(x,0) \).
This implies that:\begin{eqnarray}
\hat{\xi }_{\nu }^{(in)}(x,\omega ) & = & 
\sqrt{2\pi }\delta (\omega -\Omega _{\nu })\hat{\xi }_{\nu }^{(in)}(x)\\ \nonumber
\hat{\xi }_{\nu }^{(in)\dagger }(x,-\omega ) & = & 
\sqrt{2\pi }\delta (\omega +\Omega _{\nu })\hat{\xi }_{\nu }^{(in)\dagger }(x) \,
\, .
\end{eqnarray}

Application of the differential operator \( c^{2}\partial ^{2}_{x}+\omega ^{2} \)
to Eq. (\ref{solution}), this time keeping the matter terms, gives: 

\begin{equation}
\label{inhomog}
\partial _{x}\left( \frac{c^{2}}{n^{2}(x,\omega )}\partial _{x}\hat{\Lambda }(x,\omega )\right) +\omega ^{2}\hat{\Lambda }(x,\omega )=\partial _{x}\widehat{F}_{in}(x,\omega )\, \, ,
\end{equation}

where we now include an inhomogeneous term defined as:\begin{equation}
\label{inhom}
\widehat{F}_{in}(x,\omega )=c^{2}\sum _{\nu }\sqrt{\frac{\hbar \pi \epsilon _{0}g_{\nu }(x)}{\Omega _{\nu }}}\left( \delta (\omega -\Omega _{\nu })\hat{\xi }_{\nu }^{(in)}(x)+\delta (\omega +\Omega _{\nu })\hat{\xi }_{\nu }^{(in)\dagger }(x)\right) \, \, .
\end{equation}

This equation can be solved by means of a Green's function which satisfies:

\begin{equation}
\label{Greenf}
\partial _{x}\left( \frac{c^{2}}{n^{2}(x,\omega )}\partial _{x}G(x,x')\right) +\omega ^{2}G(x,x')=\delta (x-x')\, \, ,
\end{equation}
together with the proper boundary conditions. For the boundary conditions, we
shall choose \( G(x,x') \) to have only outgoing waves at \( x=\pm \infty  \).
 Any fields produced by these matter terms are generated in a finite region
and propagate outward, so that these boundary conditions are the appropriate
ones. The Green's function can be found by standard techniques, and is given
by:

\begin{eqnarray}
G(x,x') & = & \frac{1}{2i\omega cT(\omega )}u_{\ell }(x,\omega )u_{r}(x',\omega )
\, \, \, \, \, [x>x']\\ \nonumber
 & = & \frac{1}{2i\omega cT(\omega )}u_{r}(x,\omega )u_{\ell }(x',\omega )\, \, \, \, \, [x<x']\, \, .\label{greensol} 
\end{eqnarray}

The solution to Eq. (\ref{inhomog}) with only outgoing waves at \( x=\pm \infty  \)
, which we shall call \( \hat{\Lambda }_{s}(x,\omega ) \) is then:\begin{equation}
\label{inhomsol}
\hat{\Lambda }_{s}(x,\omega )=\int _{-L}^{L}dx'G(x,x')\partial _{x'}\widehat{F}_{in}(x',\omega )\, \, .
\end{equation}

Substitution of \( \hat{\Lambda }_{s}(x,\omega ) \) into Eq.(\ref{solution})
- the integral equation for \( \hat{\Lambda }(x,\omega ) \) - shows that it
is, as expected, a solution with \( \hat{\Lambda }_{in}(x,\omega )=0 \). A
complete solution for general \( \hat{\Lambda }_{in}(x,\omega ) \) can then
be obtained by adding to \( \hat{\Lambda }_{s} \) a solution of the homogeneous
equation with the proper \( \hat{\Lambda }_{in} \), as discussed in the previous
section. 

As has already been noted, \( \widehat{F}_{in}(x,\omega ) \) is only nonzero
if \( \omega =\pm \Omega _{\nu } \) , for some index \( \nu  \). At these
values, which correspond to the upper boundary of each band-edge, the index
of refraction vanishes inside the medium, and:

\begin{eqnarray}
T(\omega ) & = & \frac{c}{c-i\omega L}e^{-2i\omega L/c}\\ \nonumber
R(\omega ) & = & \frac{i\omega L}{c-i\omega L}e^{-2i\omega L/c}\, \, .
\label{resTR} 
\end{eqnarray}

For \( -L<x<L \) we have that:\begin{equation}
\label{usoln}
u_{\ell }(x,\omega )=u_{r}(x,\omega )=\frac{c}{c-i\omega L}e^{-i\omega L/c}\, \, .
\end{equation}

Therefore, for \( \omega =\pm \Omega _{\nu } \) , it follows that the Green's
function is constant, and \( \hat{\Lambda }_{s}(x,\omega ) \) is proportional
to:\begin{equation}
\label{grenconst}
\int _{-L}^{L}dx'\partial _{x'}\widehat{F}_{in}(x',\omega )=\widehat{F}_{in}(L,\omega )-\widehat{F}_{in}(-L,\omega )\, \, .
\end{equation}
Here, however, we must be careful since the source function \( \widehat{F}_{in}(x,\omega ) \)
is proportional to \( \sqrt{g_{\nu }(x)} \), which is discontinuous at \( x=\pm L \)
, so that the right-hand side of the above equation is not well-defined.

In order to resolve this difficulty, we must use the same technique as before,
and consider a continuously changing refractive index over a small boundary
region. That is, we suppose that \( n(x,\omega ) \) is \( 1 \) for \( x<-L-\delta  \)
and \( x>L+\delta  \), is equal to \( 0 \) for \( -L\leq x\leq L \), goes
continuously from \( 1 \) to \( 0 \) as \( x \) goes from \( -L-\delta  \)
to \( -L \), and goes continuously from \( 0 \) to \( 1 \) as \( x \) goes
from \( L \) to \( L+\delta  \). A similar behaviour is assumed for\( \sqrt{g_{\nu }(x)} \)
. We can then take the limit \( \delta \rightarrow 0 \). Let us examine what
happens in the interval between \( -L-\delta  \) and \( -L \); the interval
between \( L \) and \( L+\delta  \) is similar. 

For \( x>L+\delta  \) , the integral we wish to consider is then:
\begin{eqnarray}
\int _{-L-\delta }^{L+\delta }dx'u_{r}(x',\omega )
\partial _{x'}\widehat{F}_{in}(x',\omega ) & = & 
\int _{-L-\delta }^{-L}dx'u_{r}(x',\omega )\partial _{x'}
\widehat{F}_{in}(x',\omega )\\ \nonumber
 & + & u_{r}(L,\omega )
 \left( \widehat{F}_{in}(L,\omega )-\widehat{F}_{in}(-L,\omega )\right) \\ \nonumber
 & + & \int _{L}^{L+\delta }dx'u_{r}(x',\omega )
 \partial _{x'}\widehat{F}_{in}(x',\omega )\, \, .
\end{eqnarray}

For \( x \) in other intervals, the situation is similar. Note that \( u_{r}(x',\omega ) \)
is now a solution of the homogeneous version of Eq. (\ref{Greenf}), including
the modified index of refraction. From this equation it is relatively straightforward
to show that, for \( \delta  \) small,\begin{equation}
\label{smalldelta}
\int _{L}^{L+\delta }dx'u_{r}(x',\omega )\partial _{x'}\widehat{F}_{in}(x',\omega )\approx u_{r}(L,\omega )\left( \widehat{F}_{in}(L+\delta ,\omega )-\widehat{F}_{in}(L,\omega )\right) \, \, ,
\end{equation}

and this becomes an equality as \( \delta \rightarrow 0 \) . A similar relationship
holds for the integral from \( -L-\delta  \) to \( -L \) : 

\begin{equation}
\label{smalldelta2}
\int _{-L-\delta }^{-L}dx'u_{r}(x',\omega )\partial _{x'}\widehat{F}_{in}(x',\omega )\approx u_{r}(-L,\omega )\left( \widehat{F}_{in}(-L-\delta ,\omega )-\widehat{F}_{in}(-L,\omega )\right) \, \, .
\end{equation}

Adding these contributions up, and noting that if \( n(x,\omega )=0 \) for
\( -L\leq x\leq L \), then \( u_{r}(-L,\omega )=u_{r}(L,\omega ) \), we find
that:

\begin{equation}
\label{zero-int}
\lim _{\delta \rightarrow 0}\int _{-L-\delta }^{L+\delta }dx'u_{r}(x',\omega )\partial _{x'}\widehat{F}_{in}(x',\omega )=0\, \, .
\end{equation}

This implies that the matter operators \emph{do not contribute} to the interpolating
field solution, even at the resonance frequencies which we did not consider
in detail previously. Thus, the previous relation between the in and out operators
still holds at the resonances where \( \omega =\pm \Omega _{\nu } \) . At first
sight, this seems difficult to understand, since in general one would need to
include noise operators to conserve commutation relations, and hence unitarity.
However, this is consistent because, as can be seen from Eq.(\ref{resTR}),
we have \( |T(\omega )|^{2}+|R(\omega )|^{2}=1 \), even when \( \omega =\pm \Omega _{\nu } \)
for \( \nu =1,..N \) . In summary, we reach the somewhat surprising conclusion
that no additional noise operators are needed in the asymptotic properties of
the present model - even at resonance.

\section{Photodetection example}

Given the state of the in-field, these equations allow us to calculate the properties
of the out-field, and hence calculate observable scattering properties. In order
to see how this works let us consider an example. We shall find the probability
that a photo-detector located at \( x \), where \( x>0 \) and is far from
the medium, will fire at time \( t \). At the long times required for propagation
to this location, the fields will asymptotically become out-fields. The photo-detection
probability is therefore proportional to \begin{equation}
\label{corr}
\langle in|\hat{D}^{(-)}_{out}(x,t)\hat{D}^{(+)}_{out}(x,t)|in\rangle =\langle in|(\partial _{x}\hat{\Lambda }^{(-)}_{out}(x,t))(\partial _{x}\hat{\Lambda }^{(+)}_{out}(x,t))|in\rangle \, \, ,
\end{equation}
 where \( |in\rangle  \) is the in state, \begin{equation}
\hat{\Lambda }^{(+)}_{out}(x,t)=\int dk\sqrt{\frac{\hbar c\epsilon _{0}}{4\pi A|k|}}\hat{a}^{(out)}_{k}e^{i(kx-|k|ct)}\, \, ,
\end{equation}
 and \( \hat{\Lambda }^{(-)}_{out}(x,t)=(\hat{\Lambda }^{(+)}_{out}(x,t))^{\dagger } \).
Now let \( f(k) \) be a function which is zero if \( k<0 \) . The Fourier
transform of \( f(k) \) is closely related to the shape of the pulse which
is being sent into the medium. Define \begin{equation}
\hat{a}^{\dagger }_{in}[f]=\int dkf(k)(\hat{a}_{k}^{(in)})^{\dagger }\, \, ,
\end{equation}
 and let \begin{equation}
|in\rangle =\exp (\hat{a}_{in}[f]-\hat{a}^{\dagger }_{in}[f])|0\rangle _{in}\, \, .
\end{equation}
 This is a coherent state composed of wave packets with the intensity of the
field and the shape of the wave packet determined by \( f(k) \). For this state
the correlation function in Eq. (\ref{corr}) is given by \begin{equation}
\langle in|\hat{D}^{(-)}_{out}(x,t)\hat{D}^{(+)}_{out}(x,t)|in\rangle =\frac{\hbar c\epsilon _{0}}{4\pi A}\left| \int dkf(k)T(ck)e^{i(kx-|k|ct)}\right| ^{2}\, \, ,
\end{equation}
 where we have used Eq. (\ref{inout}) to relate the in and out operators, and
we have explicitly indicated the \( k \) dependence of the transmission coefficient.

As expected, this equation demonstrates explicitly that photodetection rates
are suppressed for frequency components that correspond to the dielectric absorption
bands, where \( T(\omega )\rightarrow 0 \). At these frequencies, the predominant
effect will be a strong reflection, with no photodetection occurring at the
detector location on the other side of the mirror.

\section{Conclusion}

We have presented an analysis of a quantized electromagnetic wave scattering
off a linear, dispersive medium of finite extent. The medium consists of harmonic
oscillators whose energy-level spacings can be chosen to match those of an actual
medium in the spirit of the classical Sellmeir expansion. What emerges is a
relation between the \emph{in} and \emph{out} fields which is most simply stated
in terms of their annihilation operators. The overall results are exact, and
simply expressed in terms of linear transmission and reflection coefficients.
The medium has both transmission and absorption bands. Since the model is a
full quantum-field version of the widely used Lorenz model that leads to the
Sellmeir expansion, it has a wide area of applicability to realistic dielectric
media with a variety of dispersion relations.

The final results are very similar to the classical expressions which relate
the amplitudes of the incoming and outgoing waves. The transmission and reflection
coefficients which one finds from the quantum and classical analyses are identical,
as they should be, given that the model is a linear one, and must reduce to
the classical theory in the correspondence limit. Expressions such as those
appearing in Eq. (\ref{inout}) are often used in quantum optics, and we believe
it is useful to see how they emerge from the underlying scattering theory.

An important feature of the theory treated here is that it has absorption bands,
without any corresponding noise terms in the field equation. This is due to
the dielectric model used here, in which the dielectric constant is always either
purely real or purely imaginary. In this type of model, all photons absorbed
are re-emitted. Thus, absorption simply results in a strong reflection, with
purely evanescent fields inside the dielectric. In a related phenomenological
treatment\cite{loud} , one finds similar behaviour: if the dielectric constant
is always either purely real or purely imaginary, it is possible to have dispersion
without any additional noise terms.

The reason for this is that the scattering terms alone are sufficient to ensure
that the input-output relations remain unitary, with no change in the commutators.
Thus, in the present model, there is no need for any additional source terms.
The theory therefore provides a justification for the use of simple input-output
relations to describe idealized dielectric or metallic mirrors, even when the
dielectric response is dispersive. However, for realistic media it is generally
the case that  absorption can also occur even in the transmission
bands. Treating this would require the use of more sophisticated models, including a complex refractive index.

\acknowledgements

This research was supported by the USA National Science Foundation under grant INT-9602515, and by the Australian Research Council.

\end{document}